\documentclass[aps,pre,showpacs,reprint,groupedaddress,amssymb]{revtex4-1}

\newcommand{\rmd}{\mathrm{d}}
\newcommand{\rmi}{\mathrm{i}}
\newcommand{\rme}{\mathrm{e}}
\newcommand{\bxi}{\bm{\xi}}

\newcommand{\bPsi}{\bm{\Psi}}
\newcommand{\bpsi}{\bm{\psi}}
\newcommand{\btau}{\bm{\tau}}

\usepackage{lmodern}
\usepackage[T1]{fontenc}
\usepackage{color}
\usepackage{amsmath}
\usepackage{graphicx}
\usepackage[dvipsnames]{xcolor}
\usepackage{bm}
\usepackage{epstopdf}
\usepackage[colorlinks,citecolor=blue]{hyperref} %for links

\begin{document}

\title{In-phase synchronization in complex oscillator networks by adaptive delayed feedback control}


\author{Viktor Novi\v{c}enko}
\email[]{viktor.novicenko@tfai.vu.lt}
\homepage[]{http://www.itpa.lt/~novicenko/}
%\thanks{}
\affiliation{Faculty of Physics, Vilnius University, Saul\.{e}tekio ave. 3, LT-10222 Vilnius, Lithuania}

\author{Irmantas Ratas}

\affiliation{Center for Physical Sciences and Technology, Saul\.{e}tekio ave. 3, LT-10222 Vilnius, Lithuania}

\begin{abstract}
In-phase synchronization is a special case of synchronous behavior when coupled oscillators have the same phases for any time moments. Such behavior appears naturally for nearly identical coupled limit-cycle oscillators when the coupling strength is greatly above the synchronization threshold. We investigate the general class of nearly identical complex oscillators connected into network in a context of a phase reduction approach. By treating each oscillator as a black-box possessing a single-input single-output, we provide a practical and simply realizable control algorithm to attain the in-phase synchrony of the network. For a general diffusive-type coupling law and any value of a coupling strength (even greatly below the synchronization threshold) the delayed feedback control with a specially adjusted time-delays can provide in-phase synchronization. Such adjustment of the delay times performed in an automatic fashion by the use of an adaptive version of the delayed feedback algorithm when time-delays become time-dependent slowly varying control parameters. Analytical results show that there are many arrangements of the time-delays for the in-phase synchronization, therefore we supplement the algorithm by an additional requirement to choose appropriate set of the time-delays, which minimize power of a control force. Performed numerical validations of the predictions highlights the usefulness of our approach.
\end{abstract}


\maketitle

\section{Introduction}

Synchronization phenomenon, in the narrow sense, can be defined as a dynamical state of oscillatory system, when two or more oscillators having different natural frequencies, due to the mutual coupling, start oscillating with the same frequency~\cite{pikov01,kura03,izhi07}. Such behavior is  referred as a \textit{frequency locking} regime~\cite{izhi07}. The special case of the frequency locking state is the in-phase synchronization appearing for nearly identical oscillators, when not only frequencies become the same, but also the phases. The in-phase synchrony occurs in many different situations. For example, it spontaneously appears in nature, like flagellar synchronization~\cite{Friedrich12,Klindt2017} and flashing of fireflies~\cite{Buck1988}, emerges in humans behavior (e.g. pedestrians on a bridge~\cite{Strogatz2005} and hand clapping~\cite{Neda2000a}), in electrochemical oscillations~\cite{Wickramasinghe2011,Fukushima2005}, coupled reaction-diffusion systems~\cite{nakao14} and is a desirable state in human-made systems, like optomechanical oscillators~\cite{Weiss2016} and coupled phase-locked loops~\cite{Pollakis2014}. Since the in-phase synchronization is simply visually perceived, it can be established with ``at home'' setup using metronomes~\cite{Pantaleone2002}. Interestingly, that historically first mention on synchrony in C.~Huygens' works was done on an anti-phase synchronization, the opposite state to the in-phase synchronization.

The huge impact for research on the network synchronization had the phase reduction technique. It enables an investigation of weakly coupled limit cycle oscillators connected into the network. Independently on complexity of the individual oscillatory unit, the phase reduction approach allow us to reduce the dynamics of oscillator into the single scalar dynamics, called phase~\cite{pikov01,kura03,izhi07}. Recent generalization of the phase reduction for systems with the time-delay~\cite{physd12,kot12} empower to deal with the oscillators described by delay-differential equations.

The time-delay plays a crucial role in algorithms devoted to control the synchronization of oscillatory networks. Mostly those algorithms require multiple delays, for example a coupling with inhomogeneous delays was used to stabilize prescribed patterns of synchrony in regular networks of coupled oscillators~\cite{Popovych2011,Kantner2015}, or to recognize arbitrary patterns in networks of excitable units~\cite{Lucken2017}. In our work the multiple delays are employed in the delay feedback control scheme.

The delay feedback algorithms are widely used in chaos control theory to stabilize unstable periodic orbit~\cite{pyr92,pyr06}, since it can be applied to situations, where the information about particular equations of the system is absent. The idea to employ the delay feedback signals for a different purpose, i.e. to control synchronization in oscillator network, seems to be promising and practical tool due to minimal required knowledge on equations describing the oscillator's dynamics. The papers~\cite{ros04,PhysRevE.70.041904} demonstrate an efficient suppression of synchronization in ensemble of globally coupled oscillators, via time-delayed mean field fed back to the system. In~\cite{Ratas2014} it is showed that, the periodically modulated version of the time-delayed feedback control, called act-and-wait algorithm, is able to desynchronize the oscillatory network. The numerical studies~\cite{Brandstetter2009,Hovel2010,hov_conf2007,Schoell2009} investigate the influence of the time-delayed control signals to the synchronization. The most of these studies were focused on the desynchronization of naturally synchronized oscillator network. In this work we focus on the opposite task, i.e. we try to synchronize the oscillator network, when it is naturally desynchronized. A precursor to this study is a work~\cite{Novicenko2015}, where the time-delayed feedback force applied to the individual oscillator demonstrate ability to do both -- to synchronize and to desynchronize the network of oscillators. As it is shown in~\cite{Novicenko2015}, for the in-phase synchronization regime the control parameters, i.e. the time delays, should be selected appropriately. In this paper our aim is to adapt an automatic adjustment of the delay times, in a similar fashion as in~\cite{Pyragas2011}. Combining both -- the phase reduction for the system with time-delay and the gradient descent method we provide practical algorithm to stabilize the in-phase synchronization in the oscillator network. The algorithm is designed in the spirit of the delayed feedback control algorithms and does not require any information on the particular system's equations.

The paper is organized as follows. Section~\ref{sect_model} is devoted for the mathematical background of the problem. In subsection~\ref{subsect_lcs_dfc} a general model of weakly coupled oscillators and a reduced phase model are introduced. In Subsection \ref{subsect_inph} the in-phase synchrony of the reduced phase model is analyzed. The main result of the paper is derived in subsection~\ref{subsect_grad}, where Eqs.~(\ref{main_eqs}) represent the algorithm of slowly varying time-delays to attain the in-phase synchronization. Since there are many configurations of the time-delays for the in-phase synchrony, an additional requirement to minimize the power of the control force is studied in subsection~\ref{subsect_pow}. In section~\ref{sect_num_sim} the validity of the proposed algorithm are demonstrated for the Stuart-Landau~\ref{subsect_ls} and FitzHugh-Nagumo~\ref{subsect_fhn} oscillators. Conclusions are presented in Section~\ref{sect_discuss}.

\section{Model description\label{sect_model}}

\subsection{Nearly identical weakly coupled limit cycle oscillators under delayed feedback control\label{subsect_lcs_dfc}}

We start from the general class of $N$ nearly identical limit cycle oscillators coupled via diffusive-type coupling law under single-input single-output control:
\begin{subequations}
\label{main_main} 
\begin{eqnarray}
\dot{\mathbf{x}}_i & = & \mathbf{f}_i\left(\mathbf{x}_i,u_i \right)+\varepsilon \sum_{j=1}^{N} a_{ij}\mathbf{G}_{ij}\left(\mathbf{x}_j,\mathbf{x}_i \right), \label{main}\\
s_i(t) & = & g\left( \mathbf{x}_i(t)\right), \label{measur}\\
u_i(t) & = & K_i\left[ s_i(t-\tau_i)-s_i(t) \right], \label{action}
\end{eqnarray}
\end{subequations}
where $\mathbf{x}_i \in \mathbb{R}^d$ is a $d$-dimensional state vector of the $i$-th oscillator, function $\mathbf{f}_i : \mathbb{R}^d \times \mathbb{R} \rightarrow \mathbb{R}^d$ defines  dynamics of the free $i$-th oscillator together with an action of the control force, $\varepsilon>0$ is a small coupling parameter, an adjacency matrix elements $ a_{ij} \geq 0$ encodes topology of the network, functions $ \mathbf{G}_{ij} : \mathbb{R}^d \times \mathbb{R}^d \rightarrow \mathbb{R}^d $ stands for the coupling law, $s_i \in \mathbb{R}$ is a value accessible for measurements, $ u_i \in \mathbb{R}$ -- action variable, $K_i$ and $ \tau_i $ are the gain and the time-delay of the $i$-th control force, respectively. Here we consider only the undirected topology, therefore $a_{ij}=a_{ji}$. To ensure the diffusive-type coupling, all functions $\mathbf{G}_{ij}\left(\mathbf{x}_j,\mathbf{x}_i \right)$ for identical input must be equal to zero, i.e. $\mathbf{G}_{ij} \left( \mathbf{x},\mathbf{x}\right)=\mathbf{0}$ for $i,j=1,2,\ldots, N$. We assume that the coupling is attractive, such that each coupling term attempts to reduce the difference between the coupled oscillators' states. To ensure the attractiveness of the coupling terms and a unique factorization of the expression $a_{ij} \mathbf{G}_{ij}(\cdot,\cdot)$, we will put a more accurate mathematical restrictions for the functions $\mathbf{G}_{ij}$ bellow Eq.~(\ref{coupl_f}). The free oscillators described by ordinary differential equations (ODEs) $\dot{\mathbf{x}}_i=\mathbf{f}_i\left(\mathbf{x}_i,0\right)$ have the stable limit cycle solutions $\bxi_i\left(t+T_i\right)=\bxi_i\left(t\right)$ where $T_i$ is a natural period of the $i$-th oscillator. Since the oscillators are nearly identical, $|\mathbf{f}_i(\mathbf{x},0)-\mathbf{f}_j(\mathbf{x},0)|\sim \varepsilon$. The difference of the natural periods of two oscillators $(T_j-T_i)\sim \varepsilon $ is a small quantity. To ensure a smallness of the control force, the delay-times are $(\tau_i-T_i)\sim \varepsilon$.

In order to derive a phase model for Eq.~(\ref{main_main}) we introduce a ``central'' oscillator determined by $\dot{\mathbf{x}}=\mathbf{f}(\mathbf{x},0)$, which has a stable limit cycle solution $\bxi(t+T)=\bxi(t)$ and a corresponding phase response curve $\mathbf{z}(t+T)=\mathbf{z}(t)$. The choice of the function $\mathbf{f}$ can be done almost freely, the only restriction is that $\left|\mathbf{f}\left(\mathbf{x},u \right)-\mathbf{f}_i\left(\mathbf{x},u \right)\right|$ should be of the order of $\varepsilon$. The phases dynamics in the rotating frame related to the ``central'' oscillator's frequency $\Omega = 2 \pi /T$ reads (for a derivation see Appendix):
\begin{equation}
\dot{\psi}_i=\omega_i^{\mathrm{eff}}+\varepsilon_i^{\mathrm{eff}} \sum_{j=1}^{N} a_{ij}h_{ij}\left(\psi_j-\psi_i\right).
\label{main_ph3}
\end{equation}
The coupling strength and frequencies in the phase model are changed by effective, due to influence of the delay feedback:
\begin{subequations}
\begin{eqnarray}
\varepsilon_i^{\mathrm{eff}} & =&\varepsilon \alpha(K_i C) , \label{eff_coup} \\
\omega_i^{\mathrm{eff}}&=&\omega_i + \Omega\frac{\tau_i-T_i}{T}\left[\alpha(K_i C)-1 \right], \label{eff_freq} 
\end{eqnarray}
\end{subequations}
where the function $\alpha(x) = \left (1+x \right)^{-1}$, the relative frequencies $\omega_i= \Omega_i - \Omega$ and the constant
\begin{equation}
C = \int \limits_{0}^{T} \left\lbrace \mathbf{z}^T(s)\cdot D_2\mathbf{f}\left(\bxi(s),0 \right) \right\rbrace \left\lbrace \left[\nabla g(\bxi(s))\right]^T \cdot \dot{\bxi}(s)\right\rbrace \rmd s.\label{cii} 
\end{equation}

The coupling function in phase model Eq.~(\ref{main_ph3}) is
\begin{equation}
h_{ij}\left(\chi \right) = \frac{1}{T}\int\limits_{0}^{2\pi} \left\lbrace\mathbf{z}^T\left(\frac{s}{\Omega}\right) \cdot \mathbf{G}_{ij}\left(\bxi\left(\frac{s+\chi}{\Omega}\right),\bxi\left(\frac{s}{\Omega}\right) \right) \right\rbrace \rmd s. \label{coupl_f}
\end{equation}
Due to the diffusive-type coupling law represented by $\mathbf{G}_{ij} \left( \mathbf{x}_j,\mathbf{x}_i\right)$, the coupling function $h_{ij}\left(\chi \right)$ also preserves this property $h_{ij}(0)=0$. Moreover, $\mathbf{G}_{ij} \left( \mathbf{x}_j,\mathbf{x}_i\right)$ should be chosen such, that derivative of the coupling function at the zero point will be positive,  $h_{ij}^{\prime}(0)=\eta_{ij}>0$. This condition guarantee the attractive coupling between oscillators. Additionally, to make the factorization of $a_{ij}\mathbf{G}_{ij}$ unique up to a constant, one should require that $\eta_{ij}=\eta$ will be the same for all couplings $\mathbf{G}_{ij}$.

The phase model Eq.~(\ref{main_ph3}) is valid only for the stable periodic orbit $\bxi(t)$. Due to action of the control force (\ref{action}), the periodic orbit can loss stability at some value of $K_i$. At the time of publication, there are no handy criteria to guarantee the stability of $\bxi(t)$. On the other hand, from a chaos control theory, a criterion which guarantees the destabilization of the periodic solution $\bxi(t)$ is known. The odd number limitation theorem~\cite{hoo12} states that the orbit $\bxi(t)$ become unstable if an inequality
\begin{equation}
K_i C < -1.\label{onl1}
\end{equation}
holds. The last inequality impose a restriction on possible values of $K_i$ in order to have the valid phase model~(\ref{main_ph3}). The sign of the constant $C$ defines the possible stability interval for the control gain $K_i$. For the positive $C$ it is $K_i \in (-1/C,\infty)$, while for negative -- $K_i \in (-\infty,-1/C)$. It is important to emphasize that, these intervals does not guarantee the stability, as the exact stability interval  depends on the functions $\mathbf{f}_i(\mathbf{x}_i,u_i)$ and $g(\mathbf{x}_i)$ and may be smaller. In subsection~\ref{subsect_ls} we demonstrate an example where the stability interval restricted only by~(\ref{onl1}), while subsection~\ref{subsect_fhn} analyze situation with the smaller stability interval.

\begin{figure} %[h!]
\centering\includegraphics[width=0.9\columnwidth]{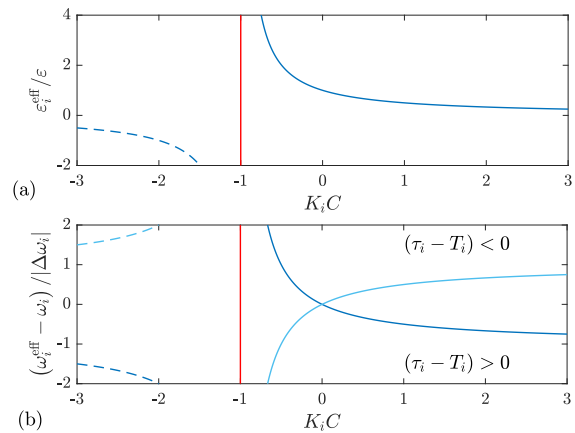}
\caption{\label{fig00}(color online) (a) Dependence of the effective coupling strength on the gain of the control force. Solid line shows potentially stable branch of the limit cycle $\bxi(t)$, while dashed line represents unstable branch. (b) Dependence of the effective frequency on the gain of the control force. The $y$-axis shows difference between the effective $\omega_i^{\mathrm{eff}}$ and the relative $\omega_i$ frequencies, normalized to quantity $|\Delta\omega_i|=\Omega\left|\tau_i-T_i\right|/T$. Dark blue (dark grey) color corresponds to positive mismatch $(\tau_i-T_i)$, while light blue (light grey) color - negative mismatch. Similar to (a), the solid and dashed lines correspond to potentially stable and unstable branches, respectively.}
\end{figure}

As one can see from the phase model (\ref{main_ph3}), the delay feedback control force changes the effective frequencies and the effective coupling strengths, but does not change the coupling function $h_{ij}(\chi)$. The effective coupling strength $\varepsilon_i^{\mathrm{eff}}$ depends on the gain of the control force $K_i$, while the effective frequency $\omega_i^{\mathrm{eff}}$ depends on two parameters: $K_i$ and a delay mismatch $\left(\tau_i-T_i\right)$. Therefore, we can control the synchronization of the network by adjusting the parameters of the control force. If inequality (\ref{onl1}) is the only restriction to the control gain, then the effective coupling strength $\varepsilon_i^{\mathrm{eff}}$ can be selected from zero to infinity, as it is demonstrated in Fig.~\ref{fig00}(a). Interestingly, that the sign of $\varepsilon_i^{\mathrm{eff}}$ can not be changed. On the other hand, the effective frequencies $\omega_i^{\mathrm{eff}}$ can be shifted from $\omega_i$ to positive or negative sides by changing the sign of the mismatch $(\tau_i-T_i)$ or the sign of $K_iC$, as one can see it from Fig.~\ref{fig00}(b).

\subsection{In-phase synchonization regime \label{subsect_inph}}

For the in-phase synchronization regime, all phases of the model (\ref{main_ph3}) will become equal $\psi_{1\,\mathrm{in}}=\psi_{2\,\mathrm{in}}=\ldots=\psi_{N\,\mathrm{in}}$. 
There always exists such set of the control parameters $\left(K_i, \tau_i \right) $ which gives a stable in-phase solution. One of the obvious example would be to fix the control parameters in such a way that all effective frequencies would vanish $\omega_i^{\mathrm{eff}}=0$. Other control parameters, that satisfy in-phase condition can be found by more detailed analysis of Eq.~(\ref{main_ph3}). For that purpose we assume that the in-phase synchronization period is $T_\mathrm{in}$  and appropriate synchronization frequency $\Omega_{\mathrm{in}}=2\pi/T_\mathrm{in}$. In the rotating frame related to $\Omega_{\mathrm{in}}$, the phases $\psi_{i\,\mathrm{in}}$ do not depend on time and are equal to the same constant:
\begin{equation}
\psi_{1\,\mathrm{in}}=\psi_{2\,\mathrm{in}}=\ldots=\psi_{N\,\mathrm{in}}=\Psi.
\label{in_ph_sync}
\end{equation}
The phases in the rotating frame related with the ``central'' oscillators frequency $\Omega$ can be transformed into the rotating frame $\Omega_{\mathrm{in}}$ by a transformation $\psi_i(t)=\psi_{i\,\mathrm{in}}(t)+\omega_{\mathrm{in}}t$ where $\omega{_\mathrm{in}}=\Omega_{\mathrm{in}}-\Omega$. Thus the dynamics of $\psi_{i\,\mathrm{in}}(t)$ is described by:
\begin{equation}
\dot{\psi}_{i \,\mathrm{in}}=\omega_i^{\mathrm{eff}}-\omega{_\mathrm{in}}+\varepsilon_i^{\mathrm{eff}} \sum_{j=1}^{N} a_{ij}h_{ij}\left(\psi_{j \,\mathrm{in}}-\psi_{i \,\mathrm{in}}\right).
\label{in_ph_model}
\end{equation}
Last equations possess the in-phase solution (\ref{in_ph_sync}), if condition
\begin{equation}
\left(\tau_i-T_i\right)\left[\alpha(K_i C)-1 \right]=T \frac{\Omega_{\mathrm{in}}-\Omega_i}{\Omega}
\label{in_ph_sync1}
\end{equation}
holds. Taking into account, that $T=T_i+O(\varepsilon)$ and $\Omega=\Omega_{\mathrm{in}}+O(\varepsilon)$, without loss of accuracy, Eq.~(\ref{in_ph_sync1}) can be rewritten as
\begin{equation}
\left(\tau_i-T_i\right)\left[1-\alpha(K_i C) \right]=\left( T_{\mathrm{in}}-T_i\right).
\label{in_ph_sync2}
\end{equation}
The last expression shows how the control parameters should be adjusted in order to attain the in-phase synchrony. Indeed, once we select the desirable $T_{\mathrm{in}}$, the r.h.s. of Eq.~(\ref{in_ph_sync2}) depends on intrinsic parameters of the system, while the left hand side of Eq.~(\ref{in_ph_sync2}) depends only on the parameters of the control force.

To proof stability of the solution~(\ref{in_ph_sync}), one needs to perturb it, $\psi_{i \,\mathrm{in}}(t)=\Psi+\delta\Psi_i(t)$, and  by the use of Eq.~(\ref{in_ph_model}) derive equations for the small disturbances $\delta\Psi_i(t)$
\begin{equation}
\delta\dot{\Psi}_i=\eta\varepsilon_i^{\mathrm{eff}}\sum_{j=1}^{N} a_{ij}\left(\delta\Psi_j-\delta\Psi_i\right),
\label{d_psi}
\end{equation}
where $\eta=h_{ij}^{\prime}(0)$. In a vector form Eq.~(\ref{d_psi}) reads
\begin{equation}
\delta\dot{\bPsi}=-\eta \mathbf{E}\mathbf{L}\delta\bPsi.
\label{d_psi1}
\end{equation}
Here $\mathbf{E}=\mathrm{diag}\left[\varepsilon_1^{\mathrm{eff}},\varepsilon_2^{\mathrm{eff}},\ldots,\varepsilon_N^{\mathrm{eff}}\right]$ is a diagonal positive-definite matrix and $\mathbf{L}=\mathbf{D}-\mathbf{A}$ is a network's Laplacian matrix combined of the adjacency $(\mathbf{A})_{ij}=a_{ij}$ and a degree  $\mathbf{D}=\mathrm{diag}\left[\sum_j a_{1j},\sum_j a_{2j},\ldots,\sum_j a_{Nj}\right]$ matrices. The solution (\ref{in_ph_sync}) is stable if the matrix $\mathbf{M}=-\eta \mathbf{E}\mathbf{L}$ does not have positive eigenvalues. The network topology described by a connected undirected graph, therefore $\mathbf{L}^T=\mathbf{L}$ is a symmetric positive semi-definite matrix with the eigenvalues $0=\lambda_1<\lambda_2\leq \cdots\leq \lambda_N$. By defining a square root of the matrix $\mathbf{E}$ as $\mathbf{E}^{1/2}$ with the entries $\left(\varepsilon_i^{\mathrm{eff}}\right)^{1/2}$ on the diagonal, one can construct a symmetric matrix $\mathbf{M}^{\prime}=-\eta \mathbf{E}^{1/2} \mathbf{L} \mathbf{E}^{1/2}$ which has the same set of the eigenvalues as the matrix $\mathbf{M}$. One can see that $\mathbf{M}^{\prime}$ is a negative semi-definite matrix, thus the in-phase solution~(\ref{in_ph_sync}) is stable. Note that (\ref{in_ph_sync}) has a neutral stability direction, since one eigenvalue of $\mathbf{M}$ is equal to $0$ and a corresponding eigenvector $\mathbf{v}=\mathbf{1}$ has all entries equal to $1$. This direction represents a shift of all phases $\psi_{i\,\mathrm{in}}$ by the same amount.

The relation (\ref{in_ph_sync2}) gives simple rules to adjust the control parameters for the in-phase synchronous regime. However, to do that one needs to know at least two things: the natural periods $T_i$ and the constant $C$ included in the expression for $\alpha(K_i C)$. In the frame of our analysis, the oscillators are the black-boxes and the only measurable quantity is the scalar signal $s_i(t)$. We assume that it is impossible to disconnect particular oscillator out of the network and measure the natural period. Therefore, our goal is to derive the algorithm to automatically adjust time-delays $\tau_i$ and the algorithm should be based only on a knowledge of $s_i(t)$.

The synchronization of the phase models is determined by two competing factors: a dissimilarity of the frequencies and the coupling strength. If the frequencies of the oscillators are not equal and the network is without control, then the in-phase synchronization can be achieved only with coupling of infinite strength, $\varepsilon \rightarrow \infty$. However, in the control case the effective coupling  $\varepsilon_i^{\mathrm{eff}}$ does not necessarily must go to infinity. Controversially, $\varepsilon_i^{\mathrm{eff}}$ can be even smaller than the natural coupling $\varepsilon$, since the feedback is able to reduce the dissimilarity of effective frequencies $\omega_i^{\mathrm{eff}}$ to the zero.

\subsection{Gradient descent method for slowly varying time-delays \label{subsect_grad}}

In this subsection our goal is to derive differential equations, which should automatically move time-delays $\tau_i(t)$ to positions, where Eq.~(\ref{in_ph_sync2}) is satisfied. Based on the ideas presented in \cite{Pyragas2011}, our main steps will be as follows: to construct a potential, which has a minimum at the in-phase synchronization regime and then allow the gradient descent algorithm to minimize the potential. To do so, we assume that initial values of the control parameters are such that the oscillator network is synchronized (in frequency locking regime) and the phases of each oscillator are close to each other. In other words, we assume that we are close to the in-phase synchronization regime. Such assumption is needed to derive analytical expressions for the potential and can be relaxed in real situations. Indeed, as we will see in section~\ref{sect_num_sim}, the network starting point can be faraway from synchronous regime, still the proposed algorithm stabilize the desirable in-phase solution. Hence we believe that the algorithm is a quite universal.

Further, we will use the phase model (\ref{main_ph3}) to find the synchronization period as well as the phases of synchronized network. Let us denote the period of the frequency locking regime as $T_\mathrm{sync}$, and the appropriate phases as $\psi_{i\, \mathrm{sync}}$. These quantities will be used in the derivation of the potential. For that purpose the phase model (\ref{main_ph3}) similarly to (\ref{in_ph_model}), can be investigated in the rotating frame related to the synchronization frequency  $\Omega_{\mathrm{sync}}=2\pi/T_\mathrm{sync}$
\begin{equation}
\dot{\psi}_{i \,\mathrm{sync}}=\omega_i^{\mathrm{eff}}-\omega{_\mathrm{sync}}+\varepsilon_i^{\mathrm{eff}} \sum_{j=1}^{N} a_{ij}h_{ij}\left(\psi_{j \,\mathrm{sync}}-\psi_{i \,\mathrm{sync}}\right),
\label{sync_ph_mod}
\end{equation}
here $\omega{_\mathrm{sync}}=\Omega_\mathrm{sync}-\Omega$ is a relative synchronization frequency. The last equations should have a stable time-independent fixed point $\bpsi_{\mathrm{sync}}(t)=\bpsi^*_{\mathrm{sync}}$.  Any difference $(\psi^*_{i \,\mathrm{sync}}-\psi^*_{j \,\mathrm{sync}})$ is small,  as we assumed that system is near in-phase synchronization. Hence we expand the coupling functions $h_{ij}(\chi)$ (\ref{coupl_f}) into Taylor series and omit the second order terms, then Eq.~(\ref{sync_ph_mod}) reads
\begin{equation}
0=\omega_i^{\mathrm{eff}}-\omega{_\mathrm{sync}}+\eta\varepsilon_i^{\mathrm{eff}} \sum_{j=1}^{N} a_{ij}\left(\psi^*_{j \,\mathrm{sync}}-\psi^*_{i \,\mathrm{sync}}\right).
\label{sync_ph_mod1}
\end{equation}
Dividing the last equations by non-zero value $\varepsilon_i^{\mathrm{eff}}$ and summing over index $i=1,2,\ldots,N$, gives
\begin{equation}
\sum_{i=1}^{N}\frac{\omega{_\mathrm{sync}}-\omega_i^{\mathrm{eff}}}{\varepsilon_i^{\mathrm{eff}}}=\eta \sum_{i,j=1}^{N} a_{ij}\left(\psi^*_{j \,\mathrm{sync}}-\psi^*_{i \,\mathrm{sync}}\right).
\label{sync_ph_mod2}
\end{equation}
The r.h.s. of Eq.~(\ref{sync_ph_mod2}) is equal to zero due to unidirected network topology. By substituting Eqs.~(\ref{eff_coup}) and (\ref{eff_freq}) into Eq.~(\ref{sync_ph_mod2}) and using the definitions of $\omega_{\mathrm{sync}}$ and $\omega_i$ we get
\begin{equation}
T\sum_{i=1}^{N}\frac{\Omega_{\mathrm{sync}}-\Omega_i}{\Omega} \left(1+K_i C \right)+\sum_{i=1}^{N}\left( \tau_i-T_i\right)K_i C=0.
\label{sync_ph_mod3}
\end{equation}
Again, one can use the fact that $\varepsilon^2$ order terms can be neglected, thus without loss of accuracy, in last expression $\Omega$ can be replaced by $\Omega_i$ and $T$ by $T_{\mathrm{sync}}$. Finally, we obtain the synchronization period:
\begin{equation}
T_{\mathrm{sync}}=\frac{\sum_{i=1}^{N}\left( T_i+K_i C \tau_i \right)}{\sum_{i=1}^{N}\left( 1+K_i C \right)}.
\label{T_sync}
\end{equation}
From this expression several insights can be done. First, if the control-free network  ($K_i=0$)  is in synchronous regime, then the synchronization period is the average of all natural periods, $T_{\mathrm{sync}}=\bar{T}=N^{-1}\sum_i T_i$. Second, if the network under control is in synchronous regime and all control gains are the same ($K_i=K$) and time-delays coincide with the natural periods ($\tau_i=T_i$), then again $T_{\mathrm{sync}}=\bar{T}$. Finally, one can show that Eq.~(\ref{T_sync}) is consistent with Eq.~(\ref{in_ph_sync2}). Indeed, Eq.~(\ref{in_ph_sync2}) gives
\begin{equation}
K_i C \tau_i=T_{\mathrm{in}}\left(1+K_i C \right)-T_i,
\label{kctau}
\end{equation}
and by inserting it into Eq.~(\ref{T_sync}) we obtain $T_{\mathrm{sync}}=T_{\mathrm{in}}$.

Next step is to obtain the phases $\psi^*_{i \,\mathrm{sync}}$. Starting from Eq.~(\ref{sync_ph_mod1}) and using similar mathematical routine as to derive $T_{\mathrm{sync}}$, one can obtain expression for the fixed point $\bpsi^*_{\mathrm{sync}}$ in a vector form:
\begin{equation}
\mathbf{L}\bpsi^*_{\mathrm{sync}}=\frac{2\pi}{\eta \varepsilon T^2}\left[T_{\mathrm{sync}} \left(\mathbf{I}+C\mathbf{K} \right) \mathbf{1} -\mathbf{T}-C\mathbf{K} \btau\right],
\label{fix}
\end{equation}
here $\mathbf{I}$ is $N\times N$ identity matrix, $\mathbf{K}=\mathrm{diag}\left[K_1,K_2,\ldots,K_N\right]$ diagonal matrix of the control gains, $\mathbf{1}$ is a vector with all entries equal to $1$, $\mathbf{T}$ -- vector of the natural periods, $\btau$ -- vector of the time-delays. The matrix $\mathbf{L}$ is singular, thus Eq.~(\ref{fix}) can have either many solutions or no solutions. Denoting $\mathbf{L}^{\dagger}$ as a Moore-Penrose pseudo-inverse of the Laplacian matrix, one can obtain that $\left(\mathbf{L}\mathbf{L}^{\dagger}\right)_{ij}=-N^{-1}+\delta_{ij}$ where $\delta_{ij}$ is a Kronecker delta. Equation (\ref{fix}) has many solutions if and only if $\mathbf{L}\mathbf{L}^{\dagger} \mathbf{b}=\mathbf{b}$, where $\mathbf{b}$ denotes the vector of the r.h.s. of Eq.~(\ref{fix}). The kernel of $\mathbf{L}\mathbf{L}^{\dagger}$ is a one dimensional space characterized by the basis vector $\mathbf{1}$, and since $\mathbf{b}$ is perpendicular to the kernel ($\mathbf{1}^T \cdot \mathbf{b}=0$), equation (\ref{fix}) has many solutions
\begin{equation}
\bpsi^*_{\mathrm{sync}}=\frac{2\pi}{\eta \varepsilon T^2}\mathbf{L}^{\dagger}\left[T_{\mathrm{sync}} C\mathbf{K}  \mathbf{1} -\mathbf{T}-C\mathbf{K} \btau\right]+\left[\mathbf{I}-\mathbf{L}^{\dagger}\mathbf{L} \right]\mathbf{w},
\label{sol}
\end{equation}
where $\mathbf{w}$ is arbitrary vector. Since $\mathbf{L}^{\dagger}\mathbf{L}=\mathbf{L}\mathbf{L}^{\dagger}$, the matrix $\left[\mathbf{I}-\mathbf{L}^{\dagger}\mathbf{L} \right]$ is a matrix where all elements are the same. As a consequence Eq.~(\ref{sol}) simplifies to
\begin{equation}
\bpsi^*_{\mathrm{sync}}=\frac{2\pi}{\eta \varepsilon T^2}\mathbf{L}^{\dagger}\left[T_{\mathrm{sync}} C\mathbf{K}  \mathbf{1} -\mathbf{T}-C\mathbf{K} \btau\right]+\mathbf{1} w,
\label{sol1}
\end{equation}
where $w$ is any scalar value. For further analysis we will need a partial derivative of $\psi^*_{i \,\mathrm{sync}}$ with respect to $\tau_j$. By using Eqs.~(\ref{T_sync}) and (\ref{sol1}) the derivative reads:
\begin{equation}
\frac{\partial\psi^*_{i \,\mathrm{sync}}}{\partial\tau_j}=\frac{2\pi K_j C}{\eta \varepsilon T^2}\left[\frac{\sum_{l=1}^{N}\left( \mathbf{L}^{\dagger}\right)_{il}K_l C}{\sum_{l=1}^{N}\left( 1+K_l C \right)}-\left( \mathbf{L}^{\dagger}\right)_{ij} \right].
\label{dp_dt}
\end{equation}
If all control gains are the same ($K_i=K$), then Eq.~(\ref{dp_dt}) reads:
\begin{equation}
\frac{\partial\psi^*_{i \,\mathrm{sync}}}{\partial\tau_j}=-\frac{2\pi KC}{\eta \varepsilon T^2}\left( \mathbf{L}^{\dagger}\right)_{ij}.
\label{dp_dt1}
\end{equation}
The synchronized phase derivative is proportional to the appropriate element of pseudo-inverse of the network's Laplacian matrix. The last expression will be used in the gradient descent method.

Now let us consider a potential:
\begin{equation}
V(t)=\frac{1}{2}\sum_{i,j=1}^{N}a_{ij}\left[s_j(t)-s_i(t) \right]^2.
\label{pot}
\end{equation}
For the identical oscillators this potential is always positive except at in-phase synchronization case. For nearly identical oscillators in general case it is not true, however further we will  expand it in the terms of $\varepsilon$, and we focus on the zero term only, which for the in-phase synchronization is equal to zero. The zero-order term $V_0(t)$ of the potential can be derived by substituting $s_j(t) \rightarrow g\left( \bxi \left(t+\psi^*_{j \,\mathrm{sync}}/\Omega_{\mathrm{sync}}\right) \right)$ into Eq.~(\ref{pot}). Additionally, one can simplify $V_0(t)$ by using an arbitrary $\Omega$ instead of $\Omega_{\mathrm{sync}}$
\begin{eqnarray}
& & V_0(t)=\frac{1}{2}\sum_{i,j=1}^{N}a_{ij} \nonumber \\
& & \times \left[g\left( \bxi \left(t+\frac{\psi^*_{j \,\mathrm{sync}}}{\Omega}\right) \right)-g\left( \bxi \left(t+\frac{\psi^*_{i \,\mathrm{sync}}}{\Omega}\right) \right) \right]^2.
\label{pot0}
\end{eqnarray}
The gradient of the potential with respect to $\tau_i$
\begin{eqnarray}
\frac{\partial V_0}{\partial \tau_i} (t) &=& \frac{T}{2\pi} \sum_{j,k=1}^{N}a_{jk}\left[s_k(t)-s_j(t) \right]\nonumber \\
& &\times \left[\dot{s}_k(t)\frac{\partial \psi^*_{k \,\mathrm{sync}}}{\partial \tau_i}-\dot{s}_j(t)\frac{\partial \psi^*_{j \,\mathrm{sync}}}{\partial \tau_i} \right].
\label{pottau}
\end{eqnarray}
By using previously derived formula (\ref{dp_dt1}), the gradients can be expressed explicitly 
\begin{eqnarray}
\frac{\partial V_0}{\partial \tau_i} (t) &=& -\frac{KC}{\eta \varepsilon T} \sum_{j,k=1}^{N}a_{jk}\left[s_k(t)-s_j(t) \right]\nonumber \\ 
& &\times \left[\dot{s}_k(t)\left( \mathbf{L}^{\dagger}\right)_{ki}-\dot{s}_j(t)\left( \mathbf{L}^{\dagger}\right)_{ji} \right].
\label{pottau1}
\end{eqnarray}

The gradient descent relaxation algorithm for the time-delays can be written as ${\dot{\tau_i}=-\beta^{\prime}\partial V_0/\partial \tau_i}$ with positive relaxation constant $\beta^{\prime}$. However, one can slightly improve the automatic adjustment of the delay-times.

Firstly, the potential (\ref{pot0}) might be equal to zero at particular time moment even, if the network is not in the in-phase synchronous state. To overcome such inconvenience and to guarantee slow variation of $\tau_i$, similarly to~\cite{PhysRevLett.100.114101}, we introduce an exponentially weighted average of the gradient~(\ref{pottau1}):
\begin{equation}
q_i (t) =  \int\limits_{t_0}^t \rme^{-\nu(t-s)} \frac{\partial V_0}{\partial \tau_i} (s) \rmd s ,
\label{av_pottau}
\end{equation}
where $t_0$ is an initial time moment of the control and $\nu^{-1}>T$ is a characteristic width of the integration window. The integral form of $q_i$ is inconvenient for simulations, thus we differentiate Eq.~(\ref{av_pottau}) in time and obtain the differential equation 
\begin{equation}
\dot{q}_i  = -\nu q_i +\frac{\partial V_0}{\partial \tau_i} (t).
\label{d_av_pottau}
\end{equation}
The last equation should be solved with an initial condition $q_i(t_0)=0$. 

Secondly, we see from Eq.~(\ref{pottau}) that the gradient requires knowledge of derivative $\dot{s}_i(t)$. To avoid direct calculation of this derivative, we introduce a new variable $p_i(t)$ governed by differential equation $\dot{p}_i  = \gamma \left( s_i-p_i \right)$. The variable $p_i(t)$ represents high-pass filter, which can be used to approximate the derivatives $\dot{s}_i(t)\approx \gamma(s_i-p_i)$, if we choose $\gamma^{-1}<T$.

Thirdly, to reduce the number of independent constants, one can renormalize the variable $q_i(t) \rightarrow q_i(t)\gamma |KC| /(\eta \varepsilon T)$ and merge together factors into one positive constant
\begin{equation}
\beta^{\prime}\frac{|KC|\gamma}{\eta \varepsilon T}=\beta >0.
\label{beta}
\end{equation}

To sum it up, the network under the delayed feedback control with adaptive time-delays is governed by
\begin{subequations}
\label{main_eqs}
\begin{eqnarray}
\dot{\mathbf{x}}_i &=& \mathbf{f}_i\left(\mathbf{x}_i,u_i \right)+\varepsilon \sum_{j=1}^{N} a_{ij}\mathbf{G}_{ij}\left(\mathbf{x}_j,\mathbf{x}_i \right), \label{main_c}\\
\dot{\tau}_i &=& -\beta q_i, \label{tau_c}\\
\dot{q}_i &=& -\nu q_i -\mathrm{sgn}(KC) \sum_{j,k=1}^{N}a_{jk}\left[s_k-s_j \right] \nonumber \\
& & \times \left[(s_k-p_k)\left( \mathbf{L}^{\dagger}\right)_{ki}-(s_j-p_j)\left( \mathbf{L}^{\dagger}\right)_{ji} \right], \label{q_c}\\
\dot{p}_i &=& \gamma \left( s_i - p_i\right), \label{p_c} \\
s_i(t) &=& g\left( \mathbf{x}_i(t)\right), \label{measur_c}\\
u_i(t) &=& K\left[ s_i(t-\tau_i(t))-s_i(t) \right], \label{action_c}
\end{eqnarray}
\end{subequations}
here $\mathrm{sgn}(\cdot)$ is a signum function. As one can see from Eq.~(\ref{q_c}), the sign of $KC$ should be guessed. In subsection~\ref{subsect_inph} we proved the stability of the in-phase regime for $\beta=0$. Due to continuity, the stability of the in-phase regime should persist for small enough $\beta$. On the other hand, too small values of $\beta$ lead to very slow approach to the in-phase synchronization solution~(\ref{in_ph_sync}). Therefore, the correct choice of $\beta$ and $\mathrm{sgn}(KC)$ is out of the scope of the proposed algorithm and should be done by a trail and error method.

\subsection{Power minimization of the control force \label{subsect_pow}}

For the fixed parameters, Eqs.~(\ref{main_eqs}) possess many in-phase solutions with different $T_{\mathrm{in}}$ and the different sets of $\tau_i$. Indeed, one can put the desirable period $T_{\mathrm{in}}$ into Eq.~(\ref{in_ph_sync2}) and obtain the set of the time-delays. Thus, the logical extension to the proposed algorithm will be a minimization of a power of the control force by appropriate choice of $\tau_i$ and $T_{\mathrm{in}}$.

For the in-phase synchronization regime the control force applied to $i$-th oscillator reads
\begin{equation}
u_i(t)=K\left[g\left( \bxi_{i \, \mathrm{in}} (t-\tau_i) \right)-g\left( \bxi_{i \, \mathrm{in}} (t) \right) \right],
\label{Fi}
\end{equation}
where $\bxi_{i \, \mathrm{in}}(t+T_{\mathrm{in}})=\bxi_{i \, \mathrm{in}}(t)$ is the periodic solution of the $i$-th oscillator, when the network of oscillators is in the in-phase synchronization state. An expansion of $u_i(t)$ in the terms of $\left(T_{\mathrm{in}}-\tau_i \right)$ gives:
\begin{eqnarray}
u_i(t)&=& K \left\lbrace \nabla g\left( \bxi\left(t \frac{\Omega_{\mathrm{in}}}{\Omega} \right) \right) \cdot \dot{\bxi}\left(t \frac{\Omega_{\mathrm{in}}}{\Omega} \right) \right\rbrace \left(T_{\mathrm{in}}-\tau_i \right) \nonumber \\
& & +O\left(\varepsilon^2\right),
\label{Fi1}
\end{eqnarray}
here we use the fact that $\bxi_{i \, \mathrm{in}}\left(t/\Omega_{\mathrm{in}} \right)=\bxi\left(t/\Omega \right)+O(\varepsilon)$. The power of the control force can be defined as the exponentially weighted average
\begin{eqnarray}
P &=& \sum_{i=1}^N \int\limits_{t_0}^t \rme^{-\nu (t-s)}  u_i^2(s) \rmd s \nonumber \\
&=& I  K^2 \sum\limits_{i=1}^N \left(T_{\mathrm{in}}-\tau_i \right)^2 +O\left( \varepsilon^3 \right),
\label{pow}
\end{eqnarray}
where $I$ is the following integral
\begin{equation}
I=\int\limits_{t_0}^{t} \rme^{-\nu (t-s)} \left\lbrace \nabla g\left( \bxi\left(s \right) \right) \cdot \dot{\bxi}\left(s \right) \right\rbrace^2 \rmd s .
\label{integ}
\end{equation}
Note, in numerical simulations $I$ can be calculated similarly to Eqs.~(\ref{av_pottau}) and (\ref{d_av_pottau}). The integral $I$ does not depend on the control parameters, thus we will focus on a normalized power 
\begin{equation}
W = \frac{C^2 P}{I} = (KC)^2 \sum\limits_{i=1}^N \left(T_{\mathrm{in}}-\tau_i \right)^2 .
\label{pow1}
\end{equation}
Intuitively the lower values of the control gain $K$ give the smaller power. However, this is not true. As we will see below, the power does not depend on the control gain.

Let us split up the periods and time-delays into ``central'' period and the $\varepsilon$ order term
\begin{subequations}
\label{t_delta_t}
\begin{eqnarray}
T_i &=& T+\delta T_i, \label{per_dper}\\
T_{\mathrm{in}} &=& T+\delta T_{\mathrm{in}}, \label{perin_dperin}\\
\tau_i &=& T+\delta \tau_i. \label{tau_dtau}
\end{eqnarray}
\end{subequations}
For the simplicity, we assume that the ``central'' period is equal to the average of the natural periods of the oscillators $T=\bar{T}$, therefore $\sum\limits_{i=1}^N \delta T_i = 0$. From Eq.~(\ref{kctau}) we have
\begin{equation}
\delta \tau_i = \frac{1+K C}{KC} \delta T_{\mathrm{in}} -\frac{\delta T_i}{KC}.
\label{dtau1}
\end{equation}
The in-phase synchronization state exists for any small value of $\delta T_{\mathrm{in}}$. By substituting Eq.~(\ref{dtau1}) into Eq.~(\ref{pow1}) one get
\begin{equation}
W = \sum\limits_{i=1}^N \left(\delta T_i- \delta T_{\mathrm{in}} \right)^2 = N \delta T_{\mathrm{in}}^2 + \sum\limits_{i=1}^N \delta T_i^2.
\label{pow2}
\end{equation}
The last expression shows that the power does not depend on the control gain and it achieves minimum for $T_{\mathrm{in}}=\bar{T}$. From Eq.~(\ref{dtau1}) one can see that for the stabilized in-phase regime any difference $\left(\tau_i-\tau_j \right)$ is exactly determined, while the absolute values $\tau_i$ are not. Thus, if we shift all time-delays by the same amount the in-phase state remains stable, but it gives different power due to $\delta T_{\mathrm{in}}$ term in Eq.~(\ref{pow2}). $W$ has parabolic dependence on $\delta T_{\mathrm{in}}$, therefore by measuring $W$ at three different points of $\delta T_{\mathrm{in}}$ one can identify the minimum of the parabola. In subsection \ref{subsect_ls} we demonstrate the minimization of the power of the control force.

\section{Numerical simulations\label{sect_num_sim}}

\begin{figure} [h!]
\centering\includegraphics[width=0.45\columnwidth]{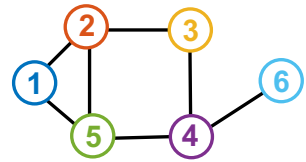}
\caption{\label{fig_topol} (color online) Topology of the oscillator network. Different colors of the nodes are used to distinguish between different oscillators in subsequent figures.}
\end{figure}

We perform numerical validation of our theory on the network of six oscillators coupled through the same function $\mathbf{G}_{ij}=\mathbf{G}$. The topology of the network is illustrated in Fig.~\ref{fig_topol}, where the connection between nodes gives $a_{ij}=1$, while $a_{ij}=0$ for unconnected nodes. We perform two different simulations: in subsection \ref{subsect_ls} we demonstrate results, when the units of network is the Sturt-Landau  oscillators and  in subsection \ref{subsect_fhn} results of network composed of FitzHugh-Nagumo neuron model is presented. The numerical integration of the state dependent DDE were implemented by standard MatLab function 'ddesd'.

\subsection{Network of Stuart-Landau oscillators\label{subsect_ls}}

As a first example, we analyze the network of the Stuart-Landau oscillators. The $i$-th oscillator's dynamics is governed by the differential equations (\ref{main_c}) where the function $\mathbf{f}_i$ reads
\begin{equation}
\mathbf{f}_i\left(\mathbf{x},u \right)=
\left[
\begin{array}{c}
x_{(1)} \left(1-x_{(1)}^2-x_{(2)}^2 \right) - \Omega_i x_{(2)} +u \\
x_{(2)} \left(1-x_{(1)}^2-x_{(2)}^2 \right) + \Omega_i x_{(1)} 
\end{array}
\right],
\label{ls_f}
\end{equation}
here $x_{(m)}$ denotes $m$-th component of the vector $\mathbf{x}$.  The coupling was chosen as follows
\begin{equation} 
\mathbf{G}(\mathbf{y},\mathbf{x})= 
\left[
\begin{array}{c}
2(y_{(1)}-x_{(1)}) \\
0 
\end{array}
\right]
.\label{ls_G}
\end{equation}
We assume that the first dynamical variable is accessible for the measurements, therefore in Eq.~(\ref{measur_c}) the function  $g(\mathbf{x})=x_{(1)}$. 

The natural frequencies are $\Omega_i=2\pi/T_i$, where the periods are distributed as $T_i=2\pi+10^{-2}\times\left[-1.2,\, 0.4,\, 0.1,\, -0.6,\, 0.3,\, 0.8 \right]$.
We chose the vector field for the ``central'' oscillator defined by Eqs.~(\ref{ls_f}) with $\Omega=1$. Due to simplicity of the Stuart-Landau oscillator one can analytically find the periodic solution $\bxi(t) = [ \cos t, \, \sin t]^T$ and the phase response curve $\mathbf{z}(t)= [ -\sin t, \, \cos t]^T$. By using Eq.~(\ref{cii}) the constant $C$ can be obtained explicitly, $C = \pi$. We check numerically that the ``central'' oscillator become unstable only if the inequality~(\ref{onl1}) holds, thus the control gain can be selected from the interval $K\in [-\pi^{-1},\infty)$. The coupling function (\ref{coupl_f}) for the phase model reads $h(\chi) = \sin(\chi)$, therefore it corresponds to Kuramoto model~\cite{kura03}.

\begin{figure} [h]
\centering\includegraphics[width=0.85\columnwidth]{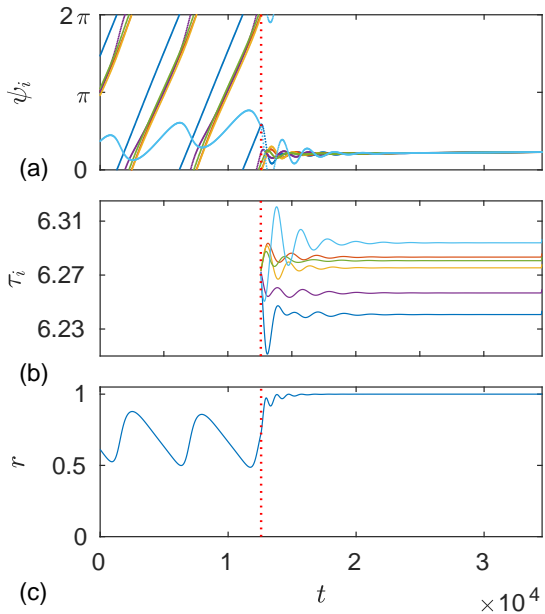}
\caption{\label{fig_ls1} (color online) Numerical simulation of the  network of Stuart-Landau oscillators. (a) The phases dynamics in the rotating frame related to the period $T_{\mathrm{in}}$; (b) Dynamics of the time delays; (c) Kuramoto order parameter. }
\end{figure}

The stabilization of the in-phase synchronization regime is demonstrated in Fig.~\ref{fig_ls1}. We choose the coupling strength $\varepsilon = 8.3\times 10^{-4}$, such that the control-free network is in desynchronized state. The network evolves uncontrolled till $t= 1.26 \times 10^4$, when the gradient descent method is turned on. The parameters of control algorithm are as follows: $K=-0.12$, $\nu=1/(10\pi)$,  $\gamma = 50/\pi$ and $\beta = 2 \times 10^{-5} $. The Fig.~\ref{fig_ls1}(a) shows phases in the rotating frame related to the settled period $T_{\mathrm{in}}$. We define the complex number $w= x_{(1)}+ \rmi x_{(2)} $ composed out of the dynamical variables of particular oscillator. The phases are estimated as follows  $\psi_i = \arg (w_i) - \Omega_{\mathrm{in}} t$. As we can see in the control free region the phases are out of consensus, while under the control all phases converge to a single constant. The Fig.~\ref{fig_ls1}(b) illustrates dynamics of the time-delays governed by Eqs.~(\ref{tau_c}). At the beginning of the control all delays are set to the same value, which after transient process settles to a fixed values. The Fig.~\ref{fig_ls1}(c) demonstrates dynamics of the Kuramoto order parameter $r = N^{-1} \left|\sum_{i} \exp(\rmi \psi_i) \right|$, which is equal to $1$ only at the in-phase synchronization regime.

It is important to emphasize, that the algorithm of the slowly varying delays is a crucial component of the control in order to achieve the synchronization. Nevertheless the control gain $K$ is such that the effective coupling strength $\varepsilon^{\mathrm{eff}}$ becomes $1.6$ times higher than the natural coupling strength $\varepsilon$, the synchronous behavior can not be achieved if all time-delays equal to the same value $\tau_i=\tau$, as it is at the beginning of control. To prove this statement, without loss of generality, one can assume that the ``central'' oscillator has the period $T=\tau$. Then, according to Eq.~(\ref{eff_freq}), the effective frequency can be written as $\omega_i^{\mathrm{eff}}=\omega_i + \frac{T_i}{T}\omega_i\left[\alpha(K)-1 \right]\approx \omega_i \alpha(K)$. Since $\varepsilon^{\mathrm{eff}}=\varepsilon \alpha(K)$, the factor $\alpha(K)$ can be eliminated from the phase model (\ref{main_ph3}) by a simple time-scaling transformation. Therefore, without the gradient descent method for the time-delays (\ref{tau_c}) not only the in-phase synchronization, but even the frequency locking regime can not be achieved.
\begin{figure} [h!]
\centering\includegraphics[width=0.85\columnwidth]{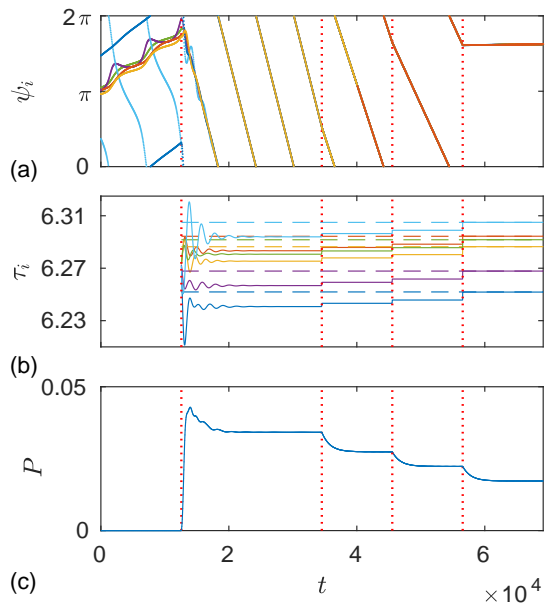}
\caption{\label{fig_ls2}(color online) The power minimization for the network of Stuart-Landau oscillators. (a) The phases dynamics in the rotating frame related to the period $\bar{T}$; (b) Dynamics of the time delays represented by solid lines and the values which minimize power depicted by the dashed lines; (c) Power of the control force.}
\end{figure}

In order to validate the ability of the power minimization of the control force, we perform additional simulations of the network of the Stuart-Landau oscillators. The results are presented in Fig.~\ref{fig_ls2}. The simulation is divided into five parts separated by the red vertical dotted lines. The first two parts coincide with the Fig.~\ref{fig_ls1}, the only difference is that in Fig.~\ref{fig_ls2}(a) the phases are estimated in the different rotating frame. This time we select the rotating frame related with the period $\bar{T}$ calculated as an average of the natural periods of the oscillators. According to Eq.~(\ref{pow2}), the minimal power is reached when $T_{\mathrm{in}}=\bar{T}$. To identify the power parabolic dependence (\ref{pow2}) on $\delta T_{\mathrm{in}}$, we shift all delays two times by the same amount (see third and fourth parts in Fig.~\ref{fig_ls2}(b)) and measure the settled powers (Fig.~\ref{fig_ls2}(c)) of the control force. The coincidence of all six phases in Fig.~\ref{fig_ls2}(a) third and fourth parts shows that such shift of the time-delays does not disrupt the in-phase synchrony as is predicted by Eq.~(\ref{dtau1}). In the last part of the simulation we set delays to the minimum of the identified parabola. The dashed lines in Fig.~\ref{fig_ls2}(b) shows analytically calculated time-delays for $\delta T_{\mathrm{in}}=0$. As one can see the analytical predictions match with the numerical simulations.

\subsection{Network of FitzHugh-Nagumo oscillators with slowly varying internal parameters\label{subsect_fhn}}

In the second example, we analyze the network of the FitzHugh-Nagumo oscillators. The dynamics of $i$-th oscillator is described by the following equations
\begin{equation}
\mathbf{f}_i\left(\mathbf{x},u \right)=
\left[
\begin{array}{c}
  x_{(1)} - x_{(1)}^3/3 -x_{(2)} + 0.5   \\
 \epsilon_i \left( x_{(1)} \left( 1+ u \right) +0.7 - 0.8 x_{(2)} \right)
\end{array}
\right].
\label{fhn_f}
\end{equation}
Here $x_{(m)}$ denotes $m$-th component of the vector $\mathbf{x}$. The oscillators differ by the parameter $\epsilon_i$, which defines the natural frequency. In experimental setup intrinsic parameters of the oscillators can vary in time due to changing external conditions or any other possible factors. The proposed control method covers such situations when the parameters vary slowly in time. To illustrate efficiency of the method, we modulated $\epsilon_i$ by harmonic functions $\epsilon_i = \epsilon + \epsilon^0_i \sin \left( w_i t + \phi_i \right)$, with different frequencies $w_i$, amplitudes $\epsilon^0_i$, and phases $\phi_i$. For this simulation, we choose non-trivial coupling law
\begin{equation} 
\mathbf{G}(\mathbf{y},\mathbf{x})= 
\left[
\begin{array}{c}
  y_{(1)} /\left( 2+ y_{(2)} \right) -  x_{(1)} / \left ( 2+ x_{(2)} \right) \\
0 
\end{array}
\right],
\label{fhn_G}
\end{equation}
and assume that the measured scalar signal $s=g(\mathbf{x})=x_{(1)}^2 + x_{(2)}$ is composed out of the first and the second variables of the oscillator.

\begin{figure} [h!]
\centering\includegraphics[width=0.85\columnwidth]{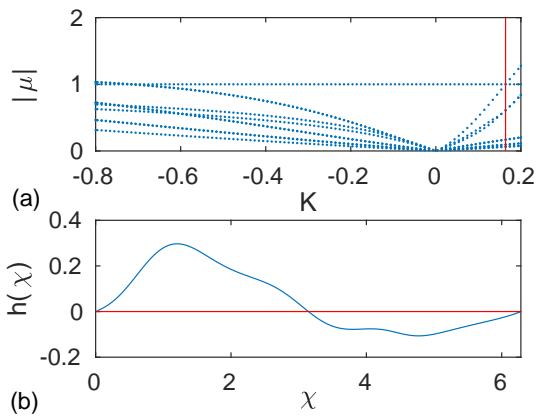}
\caption{\label{fig_fhn_h}(color online) (a) Absolute values of the first ten Floquet multipliers versus the control gain $K$. The vertical red (grey) line shows value $-C^{-1}$. (b) The coupling function $h(\chi)$ defined by Eq.~(\ref{coupl_f}) calculated for the coupling law (\ref{fhn_G}).}
\end{figure}

We chose the ``central'' oscillator having parameter $\epsilon = 0.08$. The constant $C$ calculated numerically gives $C \approx -6.1$. To check stability interval for the control gain, we calculate Floquet multipliers of the periodic solution $\bxi(t)$. According to Eq.~(\ref{onl1}), the orbit become unstable if $K>-C^{-1}$, and from Fig.~\ref{fig_fhn_h}(a) one can see that it predicts well an instability moment. However, the instability also appears for $K\lesssim -0.7$, which is not covered by Eq.~(\ref{onl1}). Figure~\ref{fig_fhn_h}(b) represents numerically calculated coupling function, which certainly differs from the harmonic function. The derivative $\eta= h^{\prime}(0)>0$ guarantees attractive coupling between the phase oscillators.

\begin{figure} [h!]
\centering\includegraphics[width=0.95\columnwidth]{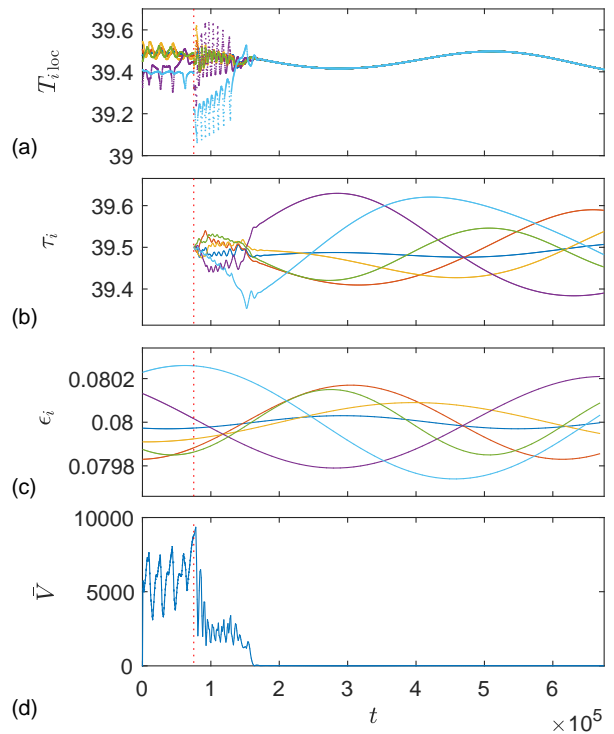}
\caption{\label{fig_fhn1}(color online) The dynamics of (a) the ``local'' periods $T_{i \, \mathrm{loc}}$; (b) the delays of the control force; (c) parameter $\epsilon_i$ that defines natural periods of FitzHugh-Nagumo oscillator; (d) the averaged potential (\ref{pot}) for the gradient descent method. The vertical red dotted line marks the moment, when control is turned on.} 
\end{figure}

\begin{figure} [h!]
\centering\includegraphics[width=0.9\columnwidth]{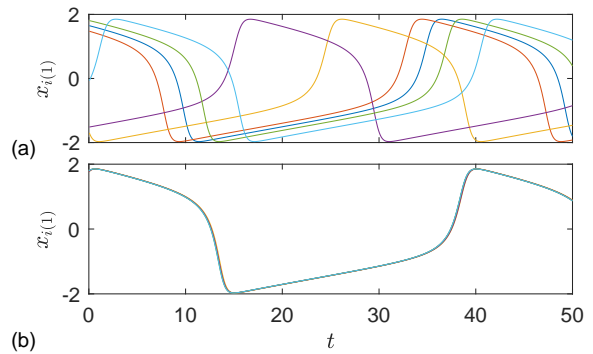}
\caption{\label{fig_fhn2}(color online) The first dynamical variable of the FitzHugh-Nagumo oscillators. The snapshots of simulations presented in figure~\ref{fig_fhn1}: (a) the control-free and (b) the controlled cases.} 
\end{figure}

The simulation results of the differential Eqs.~(\ref{main_eqs}) are demonstrated in Fig.~\ref{fig_fhn1}. In contrast to Stuart-Landau case, the dynamics of the phases $\psi_i(t)$ is difficult to extract from the dynamical variables. Therefore we calculate time distances between two neighboring maximums of the first dynamical variable and call this quantity a ``local'' period $T_{i \, \mathrm{loc}}$ (see Fig.~\ref{fig_fhn1}(a)). For the frequency locking synchronization all ``local'' periods should coincide. To confirm the in-phase synchronization, additionally we plot the potential (\ref{pot0}) in Fig.~\ref{fig_fhn1}(d). The parameters of modulation of $\epsilon_i$ are chosen as follows: $\epsilon^0_i = {[0.3,1.7,0.9,2.1,1.5,2.6] \times 10^{-4}}$, $w_i = [1.22, 1.01, 0.80, 0.80,1.36,0.80] \times 10^{-3}$, $\phi_{i}=[4.26,4.76,4.67,2.46,4.12,1.08]$. The variations of $\epsilon_i$ are showed in figure~\ref{fig_fhn1}(c). The coupling strength is set to $\varepsilon = 8 \times 10^{-4}$, control gain $K = 0.112$. Other parameters: $\beta = 3\times 10^{-7}$, $\nu = 1/\left ( 10 \bar{T} \right) \approx 2.5 \times 10^{-3}$, $\gamma = 2000/\bar{T} \approx 50.74$. The network evolves control-free till time $t_{\mathrm{on}}=7.5 \times 10^4$ (marked as red dotted line in Fig.~\ref{fig_fhn1}), when control is turned on. From Fig.~\ref{fig_fhn1}(a) one can see that before the control is turned on, the network is desynchronized as the ``local'' periods $T_{i \, \mathrm{loc}}$ are different and non-stationary. When the control is turned on, the ``local'' periods converge to a single value after the transient process. We expect that an exceptional behavior of the sixth and the fourth oscillators over transient process is related to their connectivity in the network (see Fig.~\ref{fig_topol}). At the initial stage of the control all delays are set to the same value $\tau_i(t_{\mathrm{on}})=39.5$. After the transient time, when the in-phase synchronization is reached, the time-delays still vary due to variation of $\epsilon_i$. The gradient descent method effectively decreases the exponentially weighted average of the potential, as it is shown in Fig.~\ref{fig_fhn1}(d), where it decreases 400--800 times compared with control-free case. Additionally, to ensure that in-phase synchronization is reached we present the dynamics of the first variable of the oscillators in control-free  Fig.~\ref{fig_fhn2}(a) and in controlled Fig.~\ref{fig_fhn2}(b) network.

\section{Conclusions \label{sect_discuss}}

In this paper we suggested the algorithm to achieve the in-phase synchronization state for the network of the diffusively coupled nearly identical limit cycle oscillators. The algorithm is based on  time-delayed feedback control with adaptive delay times. The method is quite universal as it does not require knowledge of the intrinsic oscillator behavior. In particular, we assume that the network units are the black-boxes having scalar output and input for measurement and for the applied control force, respectively. The control signals for each oscillator are constructed as a difference between the delayed and currently measured states multiplied by the gain factor. Such control proved to be easily realizable in experimental set-up due to its simple nature. We refer to the review paper~\cite{pyr06}, where many experimental applications are overviewed.

As we showed by Eq.~(\ref{in_ph_sync2}), the delay-feedback control is able to stabilize in-phase synchrony of the network, by proper selection of the control parameters. However, such selection requires knowledge of the intrinsic oscillator dynamics. In our framework it is impossible to disconnect particular oscillator unit out of the network. Therefore, we provide the algorithm that automatically adjusts the control parameters and stabilizes the in-phase regime. The Eq.~(\ref{in_ph_sync2}) also shows, that there exist various sets of values of control parameters that lead to in-phase synchrony. We supplement our algorithm with the minimization of total power of the control force.

Numerical demonstrations for the network of Stuart-Landau and FitzHugh-Nagumo oscillators confirm the validity of the analytically derived results. Additionally, for the case of FitzHugh-Nagumo oscillators, we show that the intrinsic parameters of the network units can slowly vary in time and the proposed algorithm still successfully manage to reach in-phase synchronization. The variation of the oscillator parameters corresponds to realistic situations in experimental set-up, where the oscillators are affected by external factors, noisy environment or have additional intrinsic slow evolution.

We expect that all listed advantages of the proposed algorithm can make it a great candidate in the experimental implementations, where the in-phase synchronization is a main objective. In particular, we expect that the algorithm can be potentially useful in a situation, where electronic components relies on a common time frame, which is attained without master clock, but due to mutual coupling between the components. For example, a global coordination between the processing cores in large multi-core systems~\cite{Pollakis2014}.

\appendix*
\section{Derivation of reduced phase model\label{subsect_ph_mod}}
Following the derivation in~\cite{Novicenko2015}, we expand the control force in the terms of $ \varepsilon $ and retain only the zeroth and the first order terms (unless otherwise stated, here and below we will always neglect higher order terms)
\begin{eqnarray}
u_i(t) &=& K_i\left[s_i(t-T_i) -s_i(t)\right] \nonumber \\
& &+K_i \dot{s}_i(t-T_i)\left(T_i-\tau_i\right)+O\left(\varepsilon^2\right).
\label{cf_expand}
\end{eqnarray}
By substituting (\ref{cf_expand}) into (\ref{main}) and expanding function $\mathbf{f}_i\left(\mathbf{x}_i,u_i \right)$ with respect to the control force, we will have
\begin{eqnarray}
\dot{\mathbf{x}}_i &=& \mathbf{f}_i\left(\mathbf{x}_i,0 \right)+D_2\mathbf{f}_i\left(\mathbf{x}_i,0 \right)K_i\left[s_i(t-T_i) -s_i(t)\right] \nonumber \\
& & +\mathbf{\Gamma}_i(\mathbf{x}_{1},\mathbf{x}_{2},...,\mathbf{x}_{N},\dot{s}_{i}(t-T_i))+O\left(\varepsilon^2\right).
\label{main_expand}
\end{eqnarray}
Here $D_2$ denote the derivation with respect to the second argument and the function
\begin{eqnarray}
& \mathbf{\Gamma}_i(\mathbf{x}_{1},\mathbf{x}_{2},...,\mathbf{x}_{N},\dot{s}_{i}(t-T_i)) =&  \nonumber \\
& D_2\mathbf{f}_i\left(\mathbf{x}_i,0 \right) K_i\dot{s}_i(t-T_i)\left(T_i-\tau_i\right) & \nonumber \\
& +\varepsilon \sum_{j=1}^{N} a_{ij}\mathbf{G}_{ij}\left(\mathbf{x}_j,\mathbf{x}_i \right),&
\label{gamma}
\end{eqnarray}
contains the first order terms with respect to $\varepsilon$. The first two terms of the right hand side (r.h.s.) of equation (\ref{main_expand}) possess the same periodic solution $\bxi_i\left(t\right)$ as the control-free oscillator. Thus one can interpret them as an oscillator without control described by delay differential equations (DDEs), while the rest terms is a small perturbation applied to it. By employing the phase reduction for the systems with time-delay~\cite{physd12} one can show that both oscillators, the ODE-oscillator and the DDE-oscillator, have the same profile of a phase response curve (PRC), the only difference is an amplitude of the PRC. The key moment here is that the second term of the r.h.s. of Eq.~(\ref{main_expand}) does not change the shape of the limit cycle, however it changes stability of the limit cycle and as a consequence the perturbation-induced phase response.

After denoting the PRC of the ODE-oscillator as $\mathbf{z}_i(t)$, the PRC of the DDE-oscillator can be expressed as $\mathbf{z}^{\mathrm{DDE}}_i(t)=\alpha\left(K_i C_i\right)\mathbf{z}_i(t)$ where the function $\alpha$ has the following form $\alpha(x) = \left( 1+x\right)^{-1}$, for more details see references~\cite{physd12,nov12}. The constant $C_i=\int_0^{T_i} c_i(s)\rmd s$ is calculated as an integral of a $T_i$-periodic auxiliary function
\begin{equation}
c_i(s) = \left\lbrace \mathbf{z}^T_i(s)\cdot D_2\mathbf{f}_i\left(\bxi_i(s),0 \right) \right\rbrace \left\lbrace \left[\nabla g(\bxi_i(s))\right]^T \cdot \dot{\bxi}_i(s)\right\rbrace.
\label{ci}
\end{equation}
Here the superscript  $\left ( \, \right)^T$ denotes transposition operation. In the following subsection we will use provided results to derive the phase model of the oscillator network~(\ref{main_expand}).

According to the phase reduction theory, the oscillators phase dynamics is described by equation
\begin{eqnarray}
& &\dot{\vartheta}_i = 1 + \left[ \mathbf{z}^{\mathrm{DDE}}_i (\vartheta_i) \right ]^T \cdot \mathbf{\Gamma}_i \left (\bxi_{1},\bxi_{2},...,\bxi_{N},\dot{s}_{i} \left (\vartheta_i (t-T_i) \right ) \right ). \nonumber \\
& &  
\label{gen_ph}
\end{eqnarray}
Here $\vartheta_i(t) \in [0, T_i)$ is the phase of the $i$-th oscillator. The first term in Eq.~(\ref{gen_ph}) represents trivial phase growth of DDE-oscillator, the second term exposes the phase change due perturbation caused by the function $\mathbf{\Gamma}_i \left (\bxi_{1},\bxi_{2},...,\bxi_{N},\dot{s}_{i} \left (\vartheta_i (t-T_i) \right ) \right )$. The states of the oscillators remain near the limit cycle, thus the periodic solutions $\bxi_i(\vartheta_i(t))$ instead of variables $\mathbf{x}_{i}(t)$ are substituted.

Note that, the function $\mathbf{\Gamma}_i \left (\bxi_{1},\bxi_{2},...,\bxi_{N},\dot{s}_{i} \left (\vartheta_i (t-T_i) \right ) \right )$ contains the delayed phases, due to term $\dot{s}_i(t-T_i)$ in Eq.~(\ref{gamma}). However, it can be avoided by neglecting the higher than $\varepsilon$-order terms, since
\begin{eqnarray}
& &\left.\dot{s}_i(t-T_i)\right|_{\bxi_i(\vartheta_i)} = \frac{\rmd}{\rmd t} \left\lbrace \left. g_i\left( \bxi_i(\vartheta)\right) \right|_{\vartheta=\vartheta_i(t-T_i)} \right\rbrace \nonumber \\
& &= \left. \left\lbrace \left[\nabla g_i(\bxi_i(\vartheta))\right]^T \cdot \dot{\bxi}_i(\vartheta)\right\rbrace \right|_{\vartheta=\vartheta_i(t-T_i)} \nonumber \\
& &= \left. \left\lbrace \left[\nabla g_i(\bxi_i(\vartheta))\right]^T \cdot \dot{\bxi}_i(\vartheta)\right\rbrace \right|_{\vartheta=\vartheta_i(t)+O(\varepsilon)} \nonumber \\
& &= \left\lbrace \left[\nabla g_i(\bxi_i(\vartheta_i(t)))\right]^T \cdot \dot{\bxi}_i(\vartheta_i(t))\right\rbrace + O(\varepsilon),
\label{s_dot}
\end{eqnarray}
and after the multiplication by $(T_i-\tau_i)$ all perturbations in $\mathbf{\Gamma}_i(\vartheta_i,\bxi_{1..N})$  will be of order of $\varepsilon$. Finally, the phase dynamics reads
\begin{eqnarray}
\label{main_ph}
\dot{\vartheta}_i &=& 1+ \alpha\left(K_i C_i\right) \left\lbrace \mathbf{z}^T_i(\vartheta_i)\cdot D_2\mathbf{f}_i\left(\bxi_i(\vartheta_i),0 \right) \right\rbrace  \\
& & \times \left\lbrace \left[\nabla g(\bxi_i(\vartheta_i))\right]^T \cdot \dot{\bxi}_i(\vartheta_i)\right\rbrace K_i \left(T_i-\tau_i\right) \nonumber \\ 
& & +\varepsilon \alpha\left(K_i C_i\right) \sum_{j=1}^{N} a_{ij}\left\lbrace\mathbf{z}^T_i(\vartheta_i) \cdot \mathbf{G}_{ij}\left(\bxi_j(\vartheta_j),\bxi_i(\vartheta_i) \right) \right\rbrace. \nonumber
\end{eqnarray}

The equation for the phase dynamics~(\ref{main_ph}) is valid only if $\bxi_i(t)$ is a stable solution of the DDE-oscillator. By the definition, $\bxi_i(t)$ is the stable solution of the ODE-oscillator. However, the second term of the r.h.s. of Eq.~(\ref{main_expand}) can destabilize it. Therefore, the stability of $\bxi_i(t)$ puts restrictions for the control gain $K_i$. At the time of publication, there are no handy criteria to guarantee the stability of $\bxi_i(t)$. On the other hand, from a chaos control theory, a criterion which guarantees the destabilization of the periodic solution $\bxi_i(t)$ is known. The odd number limitation theorem ~\cite{hoo12} states that, $\bxi_i(t)$ is an unstable solution of the DDE-oscillator, if the inequality
\begin{equation}
K_i C_i < -1,
\label{onl}
\end{equation}
holds. The last inequality impose a restriction on possible values of $K_i$ in order to have the valid phase model~(\ref{main_ph}). The sign of the constant $C_i$ defines the possible stability interval for the control gain $K_i$. For the positive $C_i$ it is $K_i \in (-1/C_i,\infty)$, while for negative -- $K_i \in (-\infty,-1/C_i)$. It is important to emphasize that, these intervals does not guarantee the stability, as the exact stability interval  depends on the functions $\mathbf{f}_i(\mathbf{x}_i,u_i)$ and $g(\mathbf{x}_i)$ and may be smaller. In subsection~\ref{subsect_ls} we demonstrate an example where the stability interval restricted only by~(\ref{onl}), while subsection~\ref{subsect_fhn} analyze situation with the smaller stability interval.

The phase model (\ref{main_ph}) can be significantly simplified. Firstly, one can see that the second term of the r.h.s. of Eq.~(\ref{main_ph}) can be written in terms of the auxiliary function $c_i$ defined by Eq.~(\ref{ci}). Secondly, the fact that the oscillators are nearly identical can be exploited. To do so, we introduce a ``central'' oscillator determined by $\dot{\mathbf{x}}=\mathbf{f}(\mathbf{x},0)$, which has a stable limit cycle solution $\bxi(t+T)=\bxi(t)$. The choice of the function $\mathbf{f}$ can be done almost freely, the only restriction is that $\left|\mathbf{f}\left(\mathbf{x},u \right)-\mathbf{f}_i\left(\mathbf{x},u \right)\right|$ should be of the order of $\varepsilon$. Thus one can write
\begin{subequations}
\label{s_main} 
\begin{eqnarray}
\bxi_i\left(s/\Omega_i\right) & =&\bxi\left(s/\Omega\right)+O(\varepsilon), \label{s_xi}\\
\mathbf{f}_i\left(\bxi_i(s/\Omega_i),0 \right)&=&\mathbf{f}\left(\bxi(s/\Omega),0 \right)+O(\varepsilon), \label{s_f} \\
\mathbf{z}_i\left(s/\Omega_i\right)&=&\mathbf{z}\left(s/\Omega\right)+O(\varepsilon) , \label{s_z} \\
c_i\left(s/\Omega_i\right)&=&c\left(s/\Omega\right) +O(\varepsilon), \label{s_ci} \\
C_i &=& C +O(\varepsilon), \label{s_c}
\end{eqnarray}
\end{subequations}
where $\Omega_i=2\pi/T_i$ is a natural frequency of the $i$-th oscillator. Using Eqs.~(\ref{s_main}) some of the indexes in (\ref{main_ph}) can be omitted:
\begin{eqnarray}
& &\dot{\vartheta}_i = 1+ \alpha\left(K_i C\right) c\left( \vartheta_i\frac{\Omega_i}{\Omega}\right)  K_i \left(T_i-\tau_i\right)+\varepsilon \alpha\left(K_i C\right) \nonumber  \\
& & \times \sum_{j=1}^{N} a_{ij}\left\lbrace\mathbf{z}^T\left(\vartheta_i\frac{\Omega_i}{\Omega}\right) \cdot \mathbf{G}_{ij}\left(\bxi\left(\vartheta_j\frac{\Omega_j}{\Omega}\right),\bxi\left(\vartheta_i\frac{\Omega_i}{\Omega}\right) \right) \right\rbrace. \nonumber \\
& & 
\label{main_ph1}
\end{eqnarray}
Accordingly, the inequality (\ref{onl}) becomes:
\begin{equation}
K_i C < -1.
\end{equation}

The phases $\vartheta_i$ grow from $0$ to $T_i$, however it is more convenient to have them growing from $0$ to $2\pi$,  when the synchronization of oscillators is investigated. Additionally, the first term on the r.h.s. of Eq.~(\ref{main_ph1}) corresponds to trivial phase growth. Therefore, we introduce new phases $\varphi_i(t)=\Omega_i\vartheta_i(t)-\Omega t$, which vary in interval $\varphi_i \in [0,2\pi)$. In terms of new variables, the phase model reads:
\begin{eqnarray}
& &\dot{\varphi}_i = \omega_i+ \Omega_i \alpha(K_i C) c\left( \frac{\varphi_i}{\Omega}+t\right) K_i\left(T_i-\tau_i\right)+\varepsilon \Omega_i\alpha(K_i C)  \nonumber \\
& &  \times \sum_{j=1}^{N} a_{ij}\left\lbrace\mathbf{z}^T\left(\frac{\varphi_i}{\Omega}+t\right)  \cdot  \mathbf{G}_{ij}\left(\bxi\left(\frac{\varphi_j}{\Omega}+t\right),\bxi\left(\frac{\varphi_i}{\Omega}+t\right) \right) \right\rbrace , \nonumber \\
& & 
\label{main_ph2}
\end{eqnarray}
here $\omega_i=\Omega_i-\Omega$ represents a relative frequency in the rotating frame related to $\Omega$. Last equations are non-autonomous, however the r.h.s. of Eq.~(\ref{main_ph2}) depends on time periodically with the period $T$. Moreover all three terms of the r.h.s. of Eq.~(\ref{main_ph2}) are proportional to small parameter $\varepsilon$. Thus one can apply averaging procedure~\cite{burd07,sand07}. Denoting averaged phases as $\psi_i(t)$, the final phase model reads:
\begin{equation}
\dot{\psi}_i=\omega_i^{\mathrm{eff}}+\varepsilon_i^{\mathrm{eff}} \sum_{j=1}^{N} a_{ij}h_{ij}\left(\psi_j-\psi_i\right),
\label{main_ph3a}
\end{equation}
here the effective coupling strength, effective frequency and coupling function read:
\begin{subequations}
\begin{eqnarray}
\varepsilon_i^{\mathrm{eff}} & =&\varepsilon \alpha(K_i C) , \label{eff_coupa} \\
\omega_i^{\mathrm{eff}}&=&\omega_i + \Omega\frac{\tau_i-T_i}{T}\left[\alpha(K_i C)-1 \right], \label{eff_freqa} \\
h_{ij}\left(\chi \right)&=&\frac{1}{T}\int\limits_{0}^{2\pi} \left\lbrace\mathbf{z}^T\left(\frac{s}{\Omega}\right) \cdot \mathbf{G}_{ij}\left(\bxi\left(\frac{s+\chi}{\Omega}\right),\bxi\left(\frac{s}{\Omega}\right) \right) \right\rbrace \rmd s. \nonumber \\
& &  \label{coupl_fa}
\end{eqnarray}
\end{subequations}
Note that the expressions (\ref{eff_freqa}) and (\ref{coupl_fa}) are written by taking into account that the frequencies $\Omega_i$ in Eq.~(\ref{main_ph2}) without loss of accuracy can be replaced by the ``central'' frequency $\Omega$.

\bibliographystyle{apsrev4-1}
\bibliography{references}

\end{document}